\documentclass[a4paper,12pt]{article}
\pdfoutput=1
\usepackage{amsmath}
\usepackage{amsfonts}
\usepackage{amssymb}
\usepackage{amstext}
\usepackage{graphicx}
\usepackage{epsfig}%

\title{\centerline \bf Intricacies of Cosmological bounce in polynomial 
metric $f(R)$ gravity for flat FLRW spacetime}
\bigskip
\author{Kaushik Bhattacharya$^\dagger$, Saikat Chakrabarty$^\ddagger$
\thanks{$^\dagger$kaushikb@iitk.ac.in, $^\ddagger$snilch@iitk.ac.in}
\\
\normalsize
Department of Physics, Indian Institute of Technology, Kanpur\\ 
\normalsize
Kanpur 208016, India
}
\begin{document}
\maketitle
\begin{abstract}
In this paper we present the techniques for computing cosmological
bounces in polynomial $f(R)$ theories, whose order is more than two, for
spatially flat FLRW spacetime. In these cases the conformally
connected Einstein frame shows up multiple scalar potentials
predicting various possibilities of cosmological evolution in the
Jordan frame where the $f(R)$ theory lives. We present a
reasonable way in which one can associate the various possible
potentials in the Einstein frame, for cubic $f(R)$ gravity, to the
cosmological development in the Jordan frame. The issue concerning the energy
conditions in $f(R)$ theories is presented. We also point out the very
important relationships between the conformal transformations
connecting the Jordan frame and the Einstein frame and the various
instabilities of $f(R)$ theory. All the calculations are done for
cubic $f(R)$ gravity but we hope the results are sufficiently general
for higher order polynomial gravity. 
\end{abstract}
\section{Introduction}
\label{intdr}

Due to advancement in technology we can presently probe most of the
information stored in the cosmic microwave background radiation (CMBR)
field. The average temperature of CMBR and the fluctuations on it
gives us a hint of the pre-CMBR phase of the universe. The data stored
in CMBR hints at the possibility that the very early phase of the
universe may have gone through an inflationary phase. The theory of
cosmic inflation \cite{Riotto:2002yw, Linde:1990ta} tries to address
various problems in cosmology, as the horizon problem, the entropy
problem, the flatness problem and others. In spite of the fact that
inflationary theories can solve most of the problems of the very early
universe they have an inherent problem, termed as the trans-Planckian
problem \cite{Brandenberger:2000wr, Martin:2000xs,
  Brandenberger:2002ty, Martin:2002kt, Brandenberger:2012aj}. To
address the difficulties of the trans-Planckian problem and the major
problem of the big-bang singularity there are proposals for an
alternative way to deal with the very early universe. The alternative
paradigm about the cosmological dynamics of the early universe deals
with the idea of a cosmological bounce. In the bouncing universe
scenario there exists a contracting phase of the universe, prior to
the time of the big-bang singularity. The contracting universe never
actually reaches the singular point but bounces to an expanding
phase. Technically the scale factor of the FLRW spacetime never
vanishes but it attains its minimum value during the bounce.
Some authors have proposed cosmological bounces in the general
relativistic framework \cite{Liu:2013kea,Piao:2003zm} and others have
used theories using modified gravity/particle physics to model such
cosmological bounces \cite{Cai:2008qw,Cai:2009rd,Cai:2012va,Cai:2013kja, Easson:2011zy, 
Bhattacharya:2013ut, Cai:2011tc, Cai:2009in, Qiu:2013eoa, Odintsov:2014gea,
Odintsov:2015zua, Odintsov:2015zza}.

Cosmological bounces can be thought of as an interesting alternative
to inflation as it resolves the horizon problem by making most part of
the observable universe to be causally connected during the
contracting phase of the universe. The perturbation length scales
corresponding to these observations became superhorizon during the
contracting phase. These perturbations later entered the causally
connected universe in the expanding phase. One may assume that far
away from the bouncing regime the dynamics of the cosmos is guided by
general relativity (GR) and only when the Ricci scalar attains a
considerable value (near bounce) the theory of gravity attains quantum
corrections \cite{Utiyama:1962sn} and an effective $f(R)$ theory of
gravity emerges \footnote{Most of the common techniques of $f(R)$
  theories which we will use in this article are appropriately summed
  up in the review articles in, Ref.~\cite{Nojiri:2010wj},
  Ref.~\cite{Sotiriou:2008rp} and Ref.~\cite{DeFelice:2010aj}.}.  One
can use $f(R)$ theory of gravity near the bouncing regime to
understand the dynamics near the bounce. It must be noted that in
$f(R)$ gravity theories one can have non-zero, and presumably, high
values of the Ricci scalar, $R$, in a primarily radiation dominated
(early) phase of the universe unlike GR, where the Ricci scalar is
zero in radiation domination. The reason being that the dynamic
equation for evolution of the metric changes in $f(R)$ theories (the
standard Einstein equation of GR is modified/altered). Due to this
fact one can indeed think of high Ricci scalar values near a radiation
dominated bouncing regime.  Some preliminary work on bouncing
cosmologies using $f(R)$ theories can be found in
Ref.~\cite{Bamba:2013fha, Novello:2008ra, Carloni:2005ii}.  In a
previous paper Ref.~\cite{Paul:2014cxa} the present authors proposed
an interesting cosmological bounce in quadratic $f(R)$ theory in
presence of hydrodynamic matter. In the previous work it was shown
that for spatially flat FLRW spacetimes one cannot get cosmological
bounces in both the Jordan frame, where the $f(R)$ theory lives, and
the conformally connected Einstein frame. In spite of this fact one
can use the Einstein frame to solve the bouncing problem.  In the
present work we extend our earlier work by touching upon the point
related to energy conditions. We show that one can always use the
Einstein frame to calculate the bouncing dynamics in general
polynomial gravity without breaking the weak energy condition and the
null energy conditions. It is also pointed out that some energy
conditions as the strong energy condition and the dominant energy
condition may be violated near the bouncing point. Actually the
cosmological bounce only happens in the Jordan frame. In this paper
when we refer to the bouncing point in the Einstein frame we mean the
cosmological coordinates in the Einstein frame which corresponds to
the cosmological variables in the Jordan frame at the time of bounce.
The energy conditions are perfectly defined in the Einstein frame
where the mathematical description of the cosmological process follows
the known route of GR. In the Jordan frame the energy conditions are
violated but this violation may not be taken very seriously as the
energy conditions are not uniquely formulated for $f(R)$ theories as
the energy momentum tensor gets contribution from curvature terms. In
this paper we explicitly deal with spatially flat FLRW spacetimes
where the dynamics of the scale factor is guided by cubic $f(R)$
theory in the Jordan frame. The interested reader can see
Ref.~\cite{Capozziello:2006dj} where the peculiarities of the two
conformal frames are discussed.

In this paper we address the important question relating to the
conformal correspondence of the Jordan frame and Einstein frame where
the form of $f(R)$ is a polynomial function, whose order is more than
two. This question is particularly important in $f(R)$ theories as
because the correspondence between the two conformal frames for such
higher order $f(R)$ theories becomes many-to-one. While there is only
one $f(R)$ theory in the Jordan frame, for a polynomial gravity theory,
there can be various different possibilities of cosmological evolution
in the Einstein frame. It appears that there are multiple Einstein
frame descriptions of a single $f(R)$ theory.  This fact emerges from
the conformal transformations connecting the two frames. For quadratic
gravity this correspondence was unique and there was no confusion in
solving the bouncing problem in the Einstein frame and then converting
the results into Jordan frame language. But as soon as one starts with
cubic gravity the problems of the conformal frames becomes
apparent. In this work we address the problem of the emergence of
multiple scalar potentials in the Einstein frame for a single
polynomial gravity theory and specify how one can identify the origin
of the multiple potentials in the Einstein frame. The discussion
regarding multiple Einstein frame description, of a unique $f(R)$
theory, is general in nature and has no particular connection to
bouncing cosmologies. We hope our results will give a new way of
interpreting polynomial $f(R)$ theories. 

In this paper we point out that there is a close relationship between
the instabilities of $f(R)$ theories and the conformal transformations
connecting the Jordan frame and the Einstein frame. It is well known
that if $f'(R) \le 0$ the gravitational theory becomes difficult to
handle as the effective gravitational constant in $f(R)$ theory
diverges or becomes negative. On the other hand if $f'(R)\le 0$ the
conformal transformations connecting the two conformal frames becomes
ill defined. Going one step further, it was shown by Dolgov and
Kawasaki in Ref.~\cite{Dolgov:2003px} that for low curvature $f(R)$
gravity the condition $f''(R) < 0$ sets up a new form of instability
where the effective gravitational interactions increases without an
upper bound. Dolgov and Kawasaki's result deals with low curvature
cases and consequently it was not clear whether their results hold for
$f(R)$ theories applied in the very early universe where the Ricci
scalar may not have a small value. In general in
Ref.~\cite{Capozziello:2006dj} the authors also proclaim that $f''(R)
> 0$ is a stability condition for their cosmological model.  In this
article we show that in the case of the very early universe, where the
gravitational dynamics is dictated by a cubic theory of gravity, the
epoch when $R$ attains the value $R_c$, where $R_c$ is the solution of
$f''(R)=0$, is not connected with other cosmological epochs.  In other
words a bouncing universe will take an infinite time to attain $R=R_c$
after the bounce. Consequently the very early universe can exist in
two complimentary branches, in one branch $f''(R)>0$ and in the other
branch $f''(R)<0$.  In the branch where $f''(R)<0$ one can have
perfectly well behaved cosmological bounces but this branch also shows
the embryo of gravitational instability, as in this branch the Ricci
scalar $R$ can become arbitrarily small and consequently
Dolgov-Kawasaki like instabilities can affect the theory in the low
curvature limit\footnote{This branch will be governed by the scalar
  potential $V_2(\phi)$ in the Einstein frame as shown later in this
  article and this branch includes the point $R=0$ where $R$ is the
  Jordan frame Ricci scalar.}. On the other hand as we are applying
$f(R)$ theory to study cosmological bounce, which can only happen in
the very early universe and at high values of the Ricci scalar, the
necessity of such a modified theory of gravity may decrease in the low
curvature limit. Consequently near the neighborhood of $R=0$ the
theory of gravity may cross-over to conventional GR.  If one uses
cubic gravity in the other branch\footnote{This branch is governed by
  the potential $V_1(\phi)$ in the Einstein frame.} then this theory
does not have any apparent instabilities but the lowest value of Ricci
scalar attainable in this branch is not zero and consequently it may
be difficult to transform this effective $f(R)$ theory into a
conventional theory of gravity as GR. The two complimentary branches
of cosmological existence meets at $R=R_c$. It is interesting to note
that the Einstein frame description of $f(R)$ gravity also breaks down
precisely at the point $R=R_c$. In this article we provide some
general analysis on the relationship of the conformal transformations
and the instabilities which plague any $f(R)$ theory of gravity. As
the instabilities are inherently related with the signs and zeros of
the first and second derivative of $f(R)$, with respect to $R$, we
will analyze the properties of these derivatives and show how
these properties affect the conformal transformations. 

The material in the present paper are organized as follows. In the
next section we briefly recapitulate the conformal relationship
between the Jordan frame and the Einstein frame for general $f(R)$
theories. In this section we present the basics of the bouncing
cosmological problem in $f(R)$ theory and opine on the very important
issues related to the signs and zeros of the first and second order
derivatives (with respect to $R$) of $f(R)$ theories and their
relationship with the conformal transformations. The second section
ends with a detailed discussion on the energy conditions in $f(R)$
theories keeping the discussion focussed around cosmological bounces.
The next section \ref{multiplicity} addresses the problem of multiple
scalar potentials which arise in the Einstein frame when $f(R)$ is a
polynomial of order more than two. In this section we choose to work
with cubic gravity where the problem of multiple scalar potentials
first shows its signature. In this same section the analysis of the
cosmological dynamics of $f(R)$ theories around $R=R_c$ is presented.
In section \ref{3pot} the ways to understand the multiple potentials
are discussed and explicit numerical solutions giving cosmological
bounces in cubic gravity are presented. This section ends with a
detailed discussion of the bouncing phenomena in cubic gravity. The
last section concludes the paper.
\section{Metric $f(R)$ gravity and its Einstein frame description}

In this section we briefly describe the relationship of the two
conformally related frames important for us. One is traditionally
referred as the Jordan frame where the metric $f(R)$ theory is
defined and the other is the Einstein frame where one can express the
gravitational evolution of the cosmological system defined in the Jordan
frame. In this article we will be particularly interested in flat FLRW 
spacetimes and consequently the discussions will be based on
equations of cosmological evolution where the curvature constant $k$
has been set to zero. The elaborate relation between these two
frames were described in an earlier paper
Ref.~\cite{Paul:2014cxa}. Here we briefly express the most important
results which connect these two frames.

Using the FLRW spacetime in terms of the scale-factor $a(t)$,
\begin{eqnarray}
ds^2 = -dt^2 + a^2(t)\left[dr^2 + r^2(d\theta^2 +
  \sin^2 \theta\,\,d\phi^2)\right]\,,
\label{frw}
\end{eqnarray}
one can write the field equations in metric $f(R)$ gravity as\cite{Sotiriou:2008rp}:
\begin{eqnarray}
3 H^2 &=&\frac{\kappa}{F(R)} \rho_{\rm eff}\,,
\label{fried}\\
3H^{2}+2\dot{H} &=&\frac{-\kappa}{F(R)} P_{\rm eff}\,,
\label{2ndeqn}
\end{eqnarray}
where $H$ is the conventional Hubble parameter defined as $H\equiv
\dot{a}/a$ and $k$ stands for the constant specifying the curvature of
the 3-dimensional spatial hypersurface. The constant $\kappa = 8\pi G$
where $G$ is the universal gravitational constant. In the above equations
\begin{eqnarray}
F(R)\equiv \frac{df(R)}{dR}\,.
\label{fp}
\end{eqnarray}
The dot specifies a derivative with respect to cosmological time $t$.
The effective energy density, $\rho_{\rm eff}$, and pressure, $P_{\rm
  eff}$, are defined as:
\begin{eqnarray}
\rho_{\rm eff} \equiv \rho + \rho _{\rm curv}\,,\,\,\,\,\,
P_{\rm eff} \equiv P + P_{\rm curv}\,,
\label{epeff}
\end{eqnarray}
where $\rho_{\rm curv}$ and $P_{\rm curv}$ are given by
\begin{eqnarray}
\rho _{\rm curv} &\equiv& \frac{RF-f}{2\kappa}-\frac{3H\dot{R}
F^{\prime}(R)}{\kappa}\,,
\label{reff}\\
P_{\rm curv} &\equiv& \frac{\dot{R}^{2}F^{\prime \prime} + 2H\dot{R}F^{\prime}
  + \ddot{R}F^{\prime} }{\kappa} - \frac{RF-f}{2\kappa}\,,
\label{peff}
\end{eqnarray}
which are curvature induced energy-density and pressure. In the above
equations the primes designate a derivative with respect to the Ricci
scalar $R$. The curvature induced thermodynamic variables exists in
absence of any hydrodynamic matter, in contrast to the conventional
$\rho$ and $P$ in
\begin{eqnarray}
T_{\mu \nu} = (\rho + P)u_\mu u_\nu + P g_{\mu\nu}\,,
\label{tmunu}
\end{eqnarray}
which has the information of hydrodynamic matter. In this article we
assume the fluid to be barotropic so that its equation of state is
\begin{eqnarray}
P=\omega \rho\,,
\label{eqns}
\end{eqnarray}
where $\omega$ is a constant and its value is zero for dust and
one-third for radiation.  It must be noted that $u_\mu$ in
Eq.~(\ref{tmunu}) is the four-velocity of a fluid element and $u_\mu
u^\mu = -1\,.$ Till now the description of the gravitational field
equation and energy momentum tensor was specified in the Jordan frame.

To understand the dynamics of $f(R)$ gravity one can recast the
problem in the Einstein frame by applying a conformal transformation
on the Jordan frame metric $g_{\mu\nu}$ as
\begin{eqnarray}
\tilde{g}_{\mu \nu}=F(R)g_{\mu \nu}\,,
\label{gtilde}
\end{eqnarray}
and simultaneously defining a new scalar field $\phi$ as
\begin{eqnarray}
\phi = \sqrt{\frac{3}{2\kappa}} \ln F(R)\,.
\label{phidef}
\end{eqnarray}
This scalar field plays an important role in the Einstein frame.  The
conformally transformed line element in the Einstein frame is
\begin{eqnarray} 
d\tilde{s}^2 = -d\tilde{t}^2 + \tilde{a}^2(\tilde{t})\left[dr^2 + r^2(d\theta^2 +
  \sin^2 \theta\,\,d\phi^2)\right]\,,
\label{dseins}
\end{eqnarray}
where the time coordinate, $\tilde{t}$, and the scale factor,
$\tilde{a}$, in the Einstein frame are related to their corresponding
Jordan frame terms via the relations $d\tilde{t}= \sqrt{F(R)} \,dt$ and
$\tilde{a}(t)=\sqrt{F(R)} \,a(t)$.  Using these transformations one can
formulate the gravitational dynamics of $f(R)$ gravity in
the Einstein frame in presence of matter and the scalar field $\phi$
acting as sources. The energy-momentum tensor in the Einstein frame, which is
related to $T^{\mu \nu}$ in the Jordan frame, turns out to be
\begin{eqnarray}
\tilde{T}_{\mu \nu} = (\tilde{\rho} +
\tilde{P})\tilde{u}_\mu \tilde{u}_\nu + \tilde{P} \tilde{g}_{\mu\nu}\,,
\label{tmnt}
\end{eqnarray}
where $\tilde{\rho}=\rho/F^2(R)$, $\tilde{P}=P/F^2(R)$ and $\tilde{u}_\mu =
\sqrt{F(R)}u_\mu$. In the Einstein frame
$\tilde{g}^{\mu\nu}\tilde{u}_\mu \tilde{u}_\nu=-1$.
Except $\tilde{T}^{\mu \nu}$, the energy-momentum tensor for the scalar field 
also acts as source of curvature in the Einstein frame and it is given as
\begin{eqnarray}
S^{\mu}_{\,\,\,\,\nu}=\partial_\alpha \phi \partial_\nu \phi\tilde{g}^{\alpha
  \mu} - \delta^\mu_\nu {\mathcal L}(\phi)\,, 
\label{smn}
\end{eqnarray}
where the scalar field Lagrangian is
\begin{eqnarray}
{\mathcal L}(\phi)= \frac12 \partial_\alpha \phi \partial_\beta\phi 
\tilde{g}^{\alpha \beta} + V(\phi)\,.
\label{philag}
\end{eqnarray}
The scalar field potential in the Einstein frame turns out to be
\begin{eqnarray}
V(\phi)=\frac{RF-f}{2\kappa F^2}\,,
\label{potphi}
\end{eqnarray}
where one has to express $R=R(\phi)$, from Eq.~(\ref{phidef}) by
inverting it, and then express $V(\phi)$ as an explicit function of
$\phi$. From the form of $S^{\mu}_{\,\,\,\,\nu}$ and the signature of
the metric used, $S^0_{\,\,\,\,0}=-\rho$ and $S^i_{\,\,\,\,i}=P$ yielding:
\begin{eqnarray}
\rho_\phi = \frac12 \left(\frac{d\phi}{d\tilde{t}}\right)^2
+ V(\phi)\,,\,\,\,\, P_\phi = \frac12 \left(\frac{d\phi}{d\tilde{t}}\right)^2
- V(\phi)\,,
\label{rps}
\end{eqnarray}
where the scalar field $\phi$ is assumed to be a function of time
only. The total energy-momentum tensor responsible for gravitational
effects in the Einstein frame is $\tilde{T}^\mu_{\,\,\,\,\nu} + S^{\mu}_{\,\,\,\,\nu}$
which is a mixed tensor with only diagonal components. 

The time coordinate, $t$, and the Hubble parameter, $H$, in the the
Jordan frame are related to the time coordinate, $\tilde{t}$, and
Hubble parameter $\tilde{H}(\equiv
\frac{1}{\tilde{a}}\frac{d\tilde{a}}{d\tilde{t}})$, in 
the Einstein frame via the relations:
\begin{eqnarray}
\tilde{t}=\int_{t_0}^t \sqrt{F(R)} dt'\,,\,\,\,\,\,\,\,
H=\sqrt{F}\left(\tilde{H}-\sqrt{\frac{\kappa}{6}}\,\frac{d\phi}{d\tilde{t}}
\right)\,.
\label{th}
\end{eqnarray}
As we will be interested mainly in bouncing cosmologies $t_0$ will be
set to zero. The instant $t_0=0$ is the bouncing time in the Jordan
frame.  The Einstein frame description of cosmology can be tackled
like FLRW spacetime in presence of a fluid and a scalar field. The
presence of the Scalar field potential $V(\phi)$ gives one a pictorial
understanding of the physical system which is lacking in the Jordan
frame. Seeing the nature of the potential and the initial conditions
of the problem one gets a hint about the possible time development of
the system.  The time evolution of the scalar field in the Einstein
frame is dictated by the equation
\begin{eqnarray}
\frac{d^2 \phi}{d\tilde{t}^2} + 3\tilde{H}\frac{d\phi}{d\tilde{t}}
+\frac{dV}{d\phi}=\sqrt{\frac{\kappa}{6}}(1-3\omega)\tilde{\rho}\,,
\label{phieqn}
\end{eqnarray}
where the equation of state of the fluid in the Jordan frame is
$P=\omega \rho$. It is interesting to note that the equation of state
of the fluid remains the same in the Einstein frame. The evolution of
the energy density in the Einstein frame is given by
\begin{eqnarray}
\frac{d\tilde{\rho}}{d\tilde{t}}+\sqrt{\frac{\kappa}{6}}(1-3\omega)\tilde{\rho}
\frac{d \phi}{d\tilde{t}} + 3\tilde{H} \tilde{\rho}(1+\omega)=0\,.
\label{rhotildeq}
\end{eqnarray}
The above two equations dictate the time evolution of $\tilde{\rho}$
and $\phi$ in the Einstein frame. To generate proper bouncing solution
from the above two equations one requires the values of only two
quantities at the bouncing time, they are $\phi(\tilde{t}_0)$,
$\left(\frac{d\phi}{d\tilde{t}}\right)_{\tilde{t}_0}$, the other
parameters are determined from these two at the bouncing
time\cite{Paul:2014cxa}. In general for a flat FLRW spacetime these
two values of the respective quantities are enough to solve the whole
system in the Einstein frame. The expression of the Hubble parameter
and its rate of change in the Einstein frame are given as
\begin{eqnarray}
\tilde{H}^2 &=& \frac{\kappa}{3}(\rho_\phi + 
\tilde{\rho})\,,
\label{htilde}\\
\frac{d\tilde{H}}{d\tilde{t}}&=&-\frac{\kappa}{2}
\left[\left(\frac{d\phi}{d\tilde{t}}\right)^2 + \frac43 \tilde{\rho}\right]\,.
\label{hprime}
\end{eqnarray}
In the next section we will use the above results to formulate a
version of energy conditions which can be applied in metric $f(R)$
gravity.

It is to be noted that the prescription for transforming from the
Jordan frame to the Einstein frame is limited and fails in a
particular case. In this paper we will show an example of this
failure. More over the two conformal frames may not be equivalent for
all cosmological scenarios, specially for bouncing cosmologies in flat
FLRW cases in the Jordan frame there corresponds no bounce in the
Einstein frame \cite{Paul:2014cxa}.
\subsection{Cosmological bounce in Jordan frame for polynomial $f(R)$}
As in this paper we will be mainly talking about cosmological bounces
so it is pertinent to give the conditions for a bounce. The conditions
for a cosmological bounce are:
\begin{eqnarray}
H(t_0)=0\,,\,\,\,\,\dot{H}(t_0) > 0\,,
\label{bconds}
\end{eqnarray}
where $H$ is the Hubble parameter in the Jordan frame where the
gravitational effects are dictated by some $f(R)$ theory of
gravity. We will assume $t=t_0$ to be the bouncing time. In this paper
most of the time we will set the origin of cosmic time at the bouncing
point and consequently $t_0=0$. From Eq.~(\ref{th}) it can be seen
that $\tilde{t}=0$ in the Einstein frame at the time of bounce, $t=0$,
in the Jordan frame.

The condition that the Hubble parameter vanishes at the bouncing time
transforms to the condition:
\begin{eqnarray}
\rho_{0} + \frac{R_{0} f^{\prime}_{0}-f_{0}}{2\kappa}=0\,.
\label{hzero}
\end{eqnarray}
for a flat FLRW spacetime. In the above equation we have used
subscripts zero to denote the values of the quantities at the bouncing
time when $t=0$. If one assumes the hydrodynamic matter in the
cosmological background satisfies the conditions\footnote{It is to be
  noted that the conditions on the energy density and pressure assumed
  here does not stem from any energy condition and one cannot strictly
  call it the weak energy condition as the authors did in
  \cite{Paul:2014cxa}. Although the conditions can be derived by
  assuming $T_{\mu \nu}u^\mu u^\nu \ge 0$ in the Jordan frame where
  $u^\mu$ is any time-like 4-vector, but then also we cannot strictly
  call this conditions the weak energy condition as $T_{\mu \nu}$ is
  not the only source of curvature, as implicitly assumed in Einstein
  gravity, as there are effective Ricci scalar dependent energy
  density and pressure.  The only way to justify the conditions is by
  assuming no exotic hydrodynamic matter during the bounce
  phase. Discussions on the energy conditions in $f(R)$ gravity will
  be presented later in this paper.}:
\begin{eqnarray}
\rho \ge 0\,,\,\,\,\,\rho + P \ge 0\,,
\label{econd}
\end{eqnarray}
then the other bouncing condition becomes
\begin{equation}
\dot{R}^{2}_{0}f^{\prime \prime \prime}_{0}+\ddot{R}_{0}f^{\prime
  \prime}_{0} < 0\,,
\label{hdotp1}
\end{equation} 
for the flat FLRW solution. In Ref.~\cite{Paul:2014cxa} it was shown
that if one takes $f(R)= R + \alpha R^2$ then the bouncing conditions
invariably constrains $\alpha < 0$ and a successful cosmological
bounce only happens in the presence of hydrodynamic matter. 

In this paper we will focus on cubic gravity where 
\begin{eqnarray}
f(R)= R + \beta R^2 + \gamma R^3\,,
\label{polinfr}
\end{eqnarray}
where $\beta$ and $\gamma$ are real numbers. Using this form of $f(R)$
one can easily show that there can be two kinds of bounces, one
without any presence of matter and the other is a bounce in presence
of hydrodynamic matter. In particular for the flat FLRW spacetime in
absence of any matter the bouncing condition becomes
\begin{eqnarray}
(R_0f'_0-f_0)=0\,,
\label{bcondwm}
\end{eqnarray}
predicting that 
\begin{eqnarray}
R_0=-\frac{\beta}{2\gamma}\,.
\label{rbn}
\end{eqnarray}
As $R_0$ is positive definite at the bouncing point for flat FLRW
spacetimes, one concludes that $\gamma$ and $\beta$ must have
different signs for cosmological bounces taking place in the absence of any
hydrodynamic matter. In the present case if one demands that
$f'(R)>0$, the coefficients $\beta$ and $\gamma$ must satisfy the
following inequality
\begin{eqnarray}
0 < \beta^2 \le 3\gamma\,.
\label{bgama}
\end{eqnarray}
From Eq.~(\ref{rbn}) it was seen that $\beta$ and $\gamma$ should have
opposite sign whereas from the above equation it can be uniquely said
that for a bouncing solution $\gamma > 0$ and $\beta < 0$. 
\subsection{A discussion on the signs and zeros of the first and
  second order derivatives of $f(R)$, with respect to $R$, and their
  relationship with the conformal transformations.}
\label{instabfr}

In this article we will be analyzing bouncing phenomena with
polynomial $f(R)$. Polynomial $f(R)$ theory themselves have some
interesting features. To discuss those features we first state the
basic issues about stability of a $f(R)$ gravity theory. From
Eqs.~(\ref{fried}) and (\ref{2ndeqn}) it is seen that $\kappa/F(R)$
acts like an effective gravitational constant in $f(R)$ theory and
consequently for $F(R)>0$ the theory is well defined. On the other
hand if $F(R)\le 0$ the theory of gravitation becomes either ill
defined or produces a negative effective gravitational constant which
makes the theory unstable. On the other hand if $F(R)>0$ but $F'(R)<0$
there can be other instabilities in the theory in the low curvature
limit. This second kind of instability is generally called a
Dolgov-Kawasaki instability \cite{Dolgov:2003px}. The authors who
first analyzed this kind of instability used a weak gravitational
field and applied $f(R)$ theory where as in our case we will never use
a weak gravitational field approximation. Consequently, we do not
expect Dolgov-Kawasaki like instability to affect the stability of the
cosmological bounce phenomena.  Unlike the Dolgov-Kawasaki case for
low curvature gravity where one obtains unstable solutions for
$F'(R)<0$, in the early universe one can have both $F'(R)>0$ as well
as $F'(R)<0$, both the branches giving rise to perfectly well behaved
bounces. It will be shown later that in higher derivative cosmology
guided by $f(R)$ theory the universe avoids to attain the value of $R$
which is a solution of $F'(R)=0$. As a consequence the two branches are
complimentary to each other. If the universe is in the $F'(R)>0$
branch it will remain so for infinite time and if it is on the other
branch it will remain there for infinite time.

In $f(R)$ theory the signs and zeros, of the first and second
derivatives of $f(R)$, plays an important part in the cosmological
dynamics in the Jordan frame and the conformal transformations
connecting these two frames. The effect of the sign of the first
derivative of $f(R)$ has been discussed in the last paragraph,
henceforth we discuss the effects of the sign of the second derivative
of $f(R)$ on the conformal transformation.  More over it will be seen
that the roots of $F'(R)=0$ plays an important role as far as the
conformal transformations are concerned.  Here we present three
general results about polynomial $f(R)$ theory and the conformal
transformations relating the Jordan frame and the Einstein
frame. Henceforth in this subsection we will assume $f(R)$ to be a
polynomial of order $n$ so that $F(R)$ is a polynomial of order $n-1$
and $F'(R)$ is a polynomial of order $n-2$. The first point about an
effective Einstein frame description of an $f(R)$ theory is as
follows. {\it Any Einstein frame description of gravity arising from
  an odd order polynomial $f(R)$ where the order of the polynomial
  $n>1$ must have multiple Einstein frame potentials.}
\begin{figure}
\begin{center}
\includegraphics[scale=1]{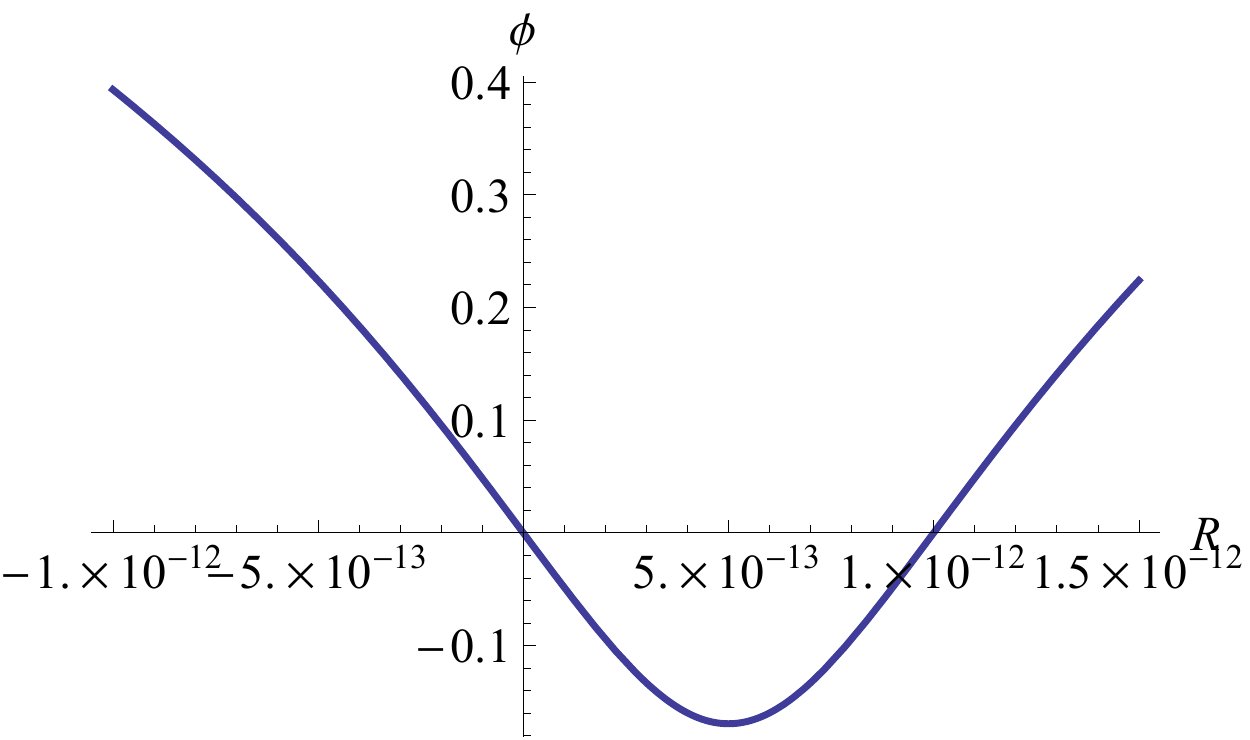} 
\caption{Einstein frame scalar field as function of Jordan frame Ricci
  scalar. We have taken $\beta=-10^{12},\,\,\gamma=\frac{2}{3}\beta^{2}$ in
  Planck units.}
\label{fig:phir}
\end{center}
\end{figure} 
To prove this statement we first notice that if one wants to write an
unique potential $V(\phi)$, as given by Eq.~(\ref{potphi}), for the
scalar field in the Einstein frame one has to invert the relation
which gives $\phi$ as a function of $R$ in Eq.~(\ref{phidef}) and
express $R=R(\phi)$. On the other hand it is seen from
Eq.~(\ref{phidef}) that $\phi(R)=\sqrt{\frac{3}{2\kappa}}\ln F(R)$ is
not uniquely invertible if $F(R)$ is a polynomial of degree greater
than or equal to two. It will be shown later that if $F(R)$ is an odd
order polynomial (hence $f(R)$ is an even order polynomial) one can
circumvent this problem related to invertibility. Only when $F(R)$ is
an even order polynomial (where $f(R)$ is an odd order polynomial) the
problem related to invertibility becomes really an acute one.  If
$F(R)$ is a polynomial of degree greater than two then the inverse
function $R=R(\phi)$ can be made single valued by partitioning it into
various branches where in each branch the relation $R=R(\phi)$ remains
a single valued monotonic increasing/decreasing function of $R$. For
all these separate branches there will be separate forms of $V(\phi)$.
If $f(R)$ is an odd order polynomial, then $F(R)$ is an even order
polynomial in $R$ and thus $F(R)$ has at least one extremum for some
finite value of $R$. Consequently, the Einstein frame description for
such an $f(R)$ theory will have multiple scalar potentials $V(\phi)$.
If $F(R)$ has $n$ extrema, then it has $n+1$ different Einstein frame
pictures separated by the points of extrema. If $f(R)$ is an even
order polynomial then $F'(R)$ is also an even order polynomial which
may not have any real zeros if the parameters of the theory are chosen
properly\footnote{In such cases there will be even number of roots of
  the equation $F'(R)=0$ and one may in principle chose half of the
  roots to be complex numbers by tuning the parameters appearing in
  the algebraic equation. The other half of them will be complex
  conjugates of the earlier roots.}. In such a case the question of
multiple potentials may be avoided. In Fig.~\ref{fig:phir} we present
the graph of the scalar field $\phi$ versus the Jordan frame Ricci
scalar $R$ for cubic $f(R)$ gravity as given in
Eq.~(\ref{polinfr}). The graph clearly shows the multi valued nature
of the inverse function. The minimum of $\phi$ appears on the right
hand side, near $R_c = 5.0 \times 10^{-13}$ where
$\phi=\phi_{c}$. Here $R_c$ is the solution of the equation $F'(R)=0$
for cubic gravity. The inverse function is invertible in the right
hand branch or the left hand branch of the minima where the branches
bifurcate at the minima. In plotting the above graph we have used
Planck units where $\beta=-10^{26}\,\,{\rm
  GeV}^{-2}=-10^{12}/M_{P}^{2}$, where $M_{P}=10^{19}$ GeV is the
approximate Planck energy. To represent the numerical values in a
brief and compact way all the dimensional quantities are scaled by
$M_{P}$ in this paper. Henceforth when we give the values of
dimensional quantities it will be assumed that we are using the Planck
units.

Our next observation regarding polynomial $f(R)$ theory is as follows.
{\it One cannot simultaneously have both $F(R)>0$ and $F'(R)>0$ for all
possible values of $R$ in a polynomial $f(R)$ theory of gravity.}
To prove the above statement one must first note that if $f(R)$ is an
even order polynomial, then $F(R)$ must be an odd order polynomial and
consequently it must have at least one real zero. As a result $F(R)>0$
does not hold for all $R$. On the other hand suppose $f(R)$ is such an
odd order polynomial so that the even order polynomial $F(R)$ has no
real roots and the condition $F(R)>0$ holds for all $R$. But then 
$F'(R)$ is an odd order polynomial and has at least one real
root and consequently one cannot have $F'(R)>0$ for all $R$.

The third observation regarding the Einstein frame description of
$f(R)$ gravity follows. {\it The Einstein frame description of $f(R)$
  gravity dynamics breaks down at the extrema of $F(R)$.}  This
observation about $f(R)$ gravity is more general in nature and also
holds for non-polynomial nature of $f(R)$.  This observation was known
previously in various forms and in Ref.~\cite{Magnano:1993bd} the
authors gave a proof of the above statement.  Here we present a
different version of the proof which is particularly suited for $f(R)$
theories governing cosmological dynamics.  To prove it one must first
note that the time evolution of the scalar field $\phi$ in the
Einstein frame is given by Eq.~(\ref{phieqn}) which contains a term
$dV/d\phi$ on the left hand side. One can calculate $dV/d\phi$ as
$$\frac{dV(\phi)}{d\phi}=\frac{dV}{dR}\frac{dR}{d\phi}\,,$$ where
Eq.~(\ref{potphi}) can be used to calculate $dV/dR$ and one can invert
Eq.~(\ref{phidef}) to calculate $dR/d\phi$. If it happens that $F(R)$
has an extremum for some value of $R$ then at that extremum point
$d\phi/dR=0$ and a result $dR/d\phi$ is divergent. In general at these
values of $R$ where $\phi$ has extremas, $dV/dR$ are well behaved.
Consequently $dV/d\phi$ becomes singular at the extremum point and the
scalar field dynamics becomes ill defined near the extremum point
making the Einstein frame description of $f(R)$ gravity
invalid.
\subsection{A brief discussion on energy conditions in $f(R)$ gravity
  and their conformal frame dependence}
\label{enrgcond}

The energy conditions on the energy-momentum tensor, for a perfect
fluid, are most well understood in Einsteins gravity where the
energy-density, $\rho$, and pressure, $P$, of the hydrodynamic fluid
acts as source of curvature in space time\footnote{In this section the
  variables $\rho$ and $P$ are used as generic symbols of
  energy-density and pressure in the Einstein frame. The same symbols
  were used earlier to specify energy-density and pressure in the
  Jordan frame. We are deliberately using these symbols as we do not
  want to increase the number of mathematical variables used in the
  paper and we think the reader can understand there role in the
  present context.}. The curvature of space time back reacts on the
source, via Einstein equation, and modifies them. As a consequence, in
cosmology guided by GR, both the metric and the fluid properties
evolve in time. This evolution by itself can produce various forms of
$\rho$ and $P$ which can produce peculiar situations where the energy
density and pressure does not behave as they should for standard
matter for which the energy conditions hold in GR. In such
circumstances one may invoke exotic matter to justify the properties
of $\rho$ and $P$. The standard energy conditions in GR can be briefly
described as \cite{poisson}:
\begin{enumerate}
\item The null energy condition (NEC) which is valid when $\rho+P \ge 0$.

\item The weak energy condition (WEC) which is valid when $\rho \ge 0$
  and $\rho+P \ge 0$. 

\item The strong energy condition (SEC) which is valid when
$\rho+ P \ge 0$ and $\rho + 3P \ge 0$.

\item The dominant energy condition (DEC) which is valid when
$\rho > 0$ and $\rho \ge |p|$.
\end{enumerate}
In the context of $f(R)$ gravity there are various complexities in
interpreting and applying energy conditions as the source of curvature
of spacetime may itself contain the Ricci scalar $R$ as seen in
Eq.~(\ref{fried}) and Eq.~(\ref{2ndeqn}). In Ref.~\cite{Santos:2007bs}
the authors tried to implement the WEC and DEC in the Jordan frame
using $\rho_{\rm eff}$ and $P_{\rm eff}$ as defined in
Eq.~(\ref{epeff}), while the NEC and SEC were derived using the
properties of Raychaudhuri's equation. In
Refs.~\cite{Capozziello:2013vna, Capozziello:2014bqa} the authors give
a much vivid discussion on the properties of energy conditions in
$f(R)$ gravity. In these references the authors note that the energy
conditions may not be simultaneously satisfied in both the Jordan and
Einstein frames. 

On the other if one started with minimally coupled scalar field in the
Einstein frame and then tried to conformally transform the theory in
the Jordan frame then matter becomes non-minimally coupled to
curvature in Jordan frame. In Ref.~\cite{Chatterjee:2012zh} the
authors tried to reformulate the NEC in the Jordan frame in such a way
that it reduces to the normal NEC in the Einstein frame under
appropriate conditions. 

In this subsection we will specifically give an example where most of
the energy conditions are violated in the Jordan frame, at the time of
bounce, and none of the energy conditions are violated in the Einstein
frame, at the corresponding time.

The structures of Eq.~(\ref{fried}) and Eq.~(\ref{2ndeqn}) and the
bouncing conditions in Eq.~(\ref{bconds}) immediately shows that for a
successful bounce in the Jordan frame 
\begin{itemize}
\item one must have at the time of bounce $(\rho_{\rm
  eff})_0=(\rho_{\rm curv})_0+\rho_0=0$ where we have assumed $\rho>0$
  in Eq.~(\ref{econd}). The subscript zero specifies the value of the
  quantities at the time of bounce. 

\item The other condition at the bouncing point is $(P_{\rm eff})_0=P_0+(P_{\rm
  curv})_0 < 0$. This implies $(P_{\rm curv})_0<0$ as we have chosen
  $P_0>0$.
\end{itemize}
These conditions show that the curvature energy is at its lowest value
during bounce, as the energy density of normal matter is maximum
during the bounce. In general after bounce $\rho$ gradually fades away
but $\rho_{\rm curv}$ increases and drives the expansion of the
universe. Consequently one can say that a cosmological bounce in
presence of matter violates the energy conditions in the Jordan frame,
if one tries to understand the energy conditions there in terms of the
effective energy density and effective pressure. On the other hand for
a bounce in the absence of hydrodynamic matter $\rho_{\rm curv} =0$
and $P_{\rm curv}<0$ at the bouncing point. Consequently in this case
also all the energy conditions are violated in the Jordan frame
if one tries to understand the energy conditions there in terms of
$\rho_{\rm eff}$ and $P_{\rm eff}$. On the other hand in the Einstein
frame the analysis of the energy conditions are different. We give the
Einstein frame analysis of the energy conditions below.

The correspondence between the $f(R)$ theory variables in the Jordan
frame in presence of a perfect fluid $T^{\mu\nu}$ and the  
conformally related Einstein frame variables was presented before.
As we have assumed normal hydrodynamic matter in the Jordan frame,
whose properties were given in Eq.~(\ref{econd}), it can easily be
shown  that in the Einstein frame one must also have 
\begin{eqnarray} 
\tilde{\rho} > 0\,,\,\,\,\,\tilde{\rho}+\tilde{P}\ge 0\,.
\label{wece}
\end{eqnarray}
The above condition does not guarantee that the various energy conditions 
will be maintained in the Einstein frame. The reason being that in
the Einstein frame one has to apply the energy conditions on
$\tilde{T}^\mu_{\,\,\,\,\nu} + S^{\mu}_{\,\,\,\,\nu}$ and consequently
the quantities which go inside the energy conditions are
$\rho_E=\tilde{\rho}+\rho_\phi$ and $P_E=\tilde{P}+P_\phi$ where
$\rho_\phi$ and $P_\phi$ are the energy density and pressure of the  
time dependent scalar field given as:
\begin{eqnarray}
\rho_\phi = \frac12 \left(\frac{d\phi}{d\tilde{t}}\right)^2
+ V(\phi)\,,\,\,\,\,
P_\phi = \frac12 \left(\frac{d\phi}{d\tilde{t}}\right)^2
- V(\phi)\,.
\label{rps1}
\end{eqnarray}
From the above equations one can notice that $\rho_\phi + P_\phi \ge
0$ always in the Einstein frame but one cannot guarantee that
$\rho_\phi>0$ in the Einstein frame and consequently the total energy
density $\rho_E$ may turn out to be negative in the Einstein
frame. This can happen because $\rho_\phi$ consists of the potential
$V(\phi)$ as given in Eq.~(\ref{potphi}) which may attain arbitrary
small negative values. If one imposes the condition that a meaningful
description of Jordan frame cosmology can only be described by its
Einstein frame description if $V(\phi)>0$ then one can be sure that
the WEC and NEC will be satisfied in the Einstein frame for most
of the cosmological processes. For bouncing cosmologies, where the
bounce happens in the Jordan frame, one cannot impose this condition
in the Einstein frame, for polynomial $f(R)$ gravity, as in these
cases the Einstein frame potential generally is always negative for
some values of $\phi$. In spite of this fact one can work out the
cosmological bounce problem in $f(R)$ gravity using the Einstein
frame, without breaking any energy conditions at the bouncing
point. One can always work in the Einstein frame in such a way that
the cosmological system respects all the energy conditions at the
``bouncing point'' (which corresponds to the actual bouncing point in
the Jordan frame), but slightly away from the bouncing point some of
the energy conditions, as the SEC or the DEC may be violated.

To elaborate the above mentioned points we specifically show the
various kinds of possible bounces in the Jordan frame and their
Einstein frame analogue. First let us concentrate bounce in the
presence of matter. From Eq.~(\ref{htilde}) one can see that
at the time of bounce, one must have
\begin{eqnarray}
V(\phi)=-\tilde{\rho}\,.
\label{bpotc}
\end{eqnarray}
As $\rho>0$ we have $\tilde{\rho}>0$ in the Einstein frame and
consequently at the bouncing time, $V(\phi)<0$ in the Einstein
frame. If near the bounce point one can ensure
$\tilde{\rho}+\frac12(\frac{d\phi}{d\tilde{t}})^2 >|V(\phi)|$ then
$\rho_E>0$, and as a result the energy conditions will be maintained in the
Einstein frame and a successful bounce can happen in the Jordan frame
where the energy conditions, interpreted in terms of the effective
thermodynamic variables, are violated.

For bounce in the absence of of matter, where $\rho=\tilde{\rho}=0$,
which is a pure curvature driven bounce we have from Eq.(\ref{bpotc}),
the potential at the time of bounce as $V(\phi)=0$ \footnote{Although
  a cosmological bounce, in the absence of any hydrodynamic matter, in
  the Jordan frame requires $V(\phi)=0$ but the vanishing of the
  Einstein frame potential does not necessarily imply a cosmological
  bounce, in the absence of hydrodynamic matter, in the Jordan frame
  in general.}. In this case at the time of bounce, in general
$\phi(0)\ne 0$ in the Einstein frame.  As bounce in absence of matter
requires $V(\phi)=0$ at the time of bounce, one can note that the
equation of state for such a scalar field in the Einstein frame will
be
\begin{eqnarray}
\omega_\phi = \frac{\frac12 \phi'^2(0) - V(\phi)}{\frac12 \phi'^2(0) + V(\phi)}
\label{omegphi}
\end{eqnarray}
which implies 
$$\omega_\phi(0) \to 1\,,$$ which implies a kinetic energy driven
free scalar field in the Einstein frame. This result is in general true for
all matter less bounces, which can be tackled in the Einstein frame,
for spatially flat FLRW spacetimes. More over as in this case
$V(\phi)=0$ at the time of bounce $\tilde{t}=0$,  
$$\rho_\phi = \frac12 \phi'^2(0)\,.$$ If one puts the condition that
$\phi'(0)=0$ then it should have been a symmetric bounce in the Jordan
frame but the peculiarity of cosmological bounces, in absence of
hydrodynamic matter, makes these symmetrical bounces in the Jordan
frame very rare. The reason being that if $\phi'(0)=0$ then $\rho_\phi
=\frac12 \phi'^2(0) + V(\phi)<0$ just near the bounce point as there
$\phi'(0) \to 0$ and $V(\phi) < 0$ breaking the energy conditions in
the Einstein frame, which makes the dynamic development of the system
impossible. Consequently any pure curvature driven bounce, where
energy conditions are maintained in the Einstein frame, mostly are
asymmetric in nature in the the Jordan frame. 

From the expressions of $\rho_\phi$ and $P_\phi$ it is clear that
$P_\phi > \rho_\phi$ if $V(\phi)<0$ and consequently DEC may be broken
near the bounce point, in the Einstein frame for pure curvature driven
bounce. On the other hand $P_\phi$ may be negative if $V(\phi)>0$ and
consequently in such regions SEC may be violated. It must be noted
that the NEC and WEC are always maintained in the Einstein frame and
none of the violations, of the other energy conditions, happen at the
bouncing point, but may be violated near to it.

The above examples show that one can have perfect cosmological bounces
in the Jordan frame, for the spatially flat FLRW spacetimes, for which
the analogous Einstein frame descriptions may not break any of the
energy conditions at the bouncing point. The energy conditions in the
Einstein frame may dictate the kind of bounce in the Jordan frame.
\section{The issue about multiple scalar potentials in the Einstein
frame}
\label{multiplicity}
In this section we present discussions on the most difficult issue
regarding the Einstein frame description of polynomial $f(R)$
theories. The issue is related with the emergence of multiple scalar
potentials $V(\phi)$ in the Einstein frame for a single $f(R)$
theory. To track the dynamics of $f(R)$ gravity in the Einstein frame
one has to solve the field equation for $\phi$ which requires the
knowledge of $V(\phi)$. From Eq.~(\ref{phieqn}) it is seen that if one
can construct a potential $V(\phi)$ in the Einstein frame using
Eq.~(\ref{potphi}) then one can uniquely determine the behavior of
$\phi$ in the Einstein frame and once this is done one can transform
the results back to the Jordan frame. This program becomes difficult
if one is unable to find out an unique scalar potential in the
Einstein frame corresponding to a unique $f(R)$ theory in the Jordan
frame. In this section we will first discuss how multiple potentials
arise, in the Einstein frame, in the case of cubic $f(R)$ theory of
gravity.  After pointing out the difficulty posed by this multiple
Einstein frame description of polynomial $f(R)$ theories we will try
to address how these difficulties can be addressed.

To illustrate the difficulties inherently related to polynomial $f(R)$
theories we study the particular case of cubic gravity . In this case
we have
\begin{eqnarray}
f(R)=R+ \beta R^2 + \gamma R^3\,,
\nonumber
\end{eqnarray}
as given in Eq.~(\ref{polinfr}).  In this case one can easily check,
from the relation
$$\phi = \sqrt{\frac{3}{2\kappa}} \ln F(R)\,,$$
that 
$$R^2 + \frac{2\beta}{3\gamma}R +
\frac{1}{3\gamma}\left(1-e^{\sqrt{2\kappa/3}\,\phi}
\right)=0\,,$$
gives $R$ in terms of $\phi$. This relation is not linear and yields
two roots as
\begin{eqnarray}
R_{1,2}=-\left(\frac{\beta}{3\gamma}\right) \pm
\sqrt{\left(\frac{\beta}{3\gamma}\right)^2 - 
\frac{1}{3\gamma}\left(1-e^{\sqrt{2\kappa/3}\,\phi}\right)}\,,
\label{r12}
\end{eqnarray}
which shows where $R_1 > R_2$ where the plus sign goes with the first
branch.  For a cosmological bounce in the Jordan frame it was
discussed before that $\gamma>0$ and $\beta<0$.  The above results
show that $R=R(\phi)$ is not a single valued function. One can tackle
this problem by using two branches for $R$ where we have for branch
one $R_1=R_1(\phi)$ and for the second branch $R_2=R_2(\phi)$. Each of
these branches are single valued functions of $\phi$. It is
interesting to note that on the first branch $f''(R_1)=F'(R_1)\ge 0$
and for the second branch $f''(R_2)=F'(R_2) \le 0$. These two branches
meet at a particular $\phi$ called $\phi_c$ whose value will be
deduced shortly. These two branches gives two different possible
values of $V(\phi)$ in the Einstein frame if one uses
Eq.~(\ref{potphi}) to express the form of the potential in the
Einstein frame.
\begin{figure}
\begin{center}
\includegraphics[scale=.75]{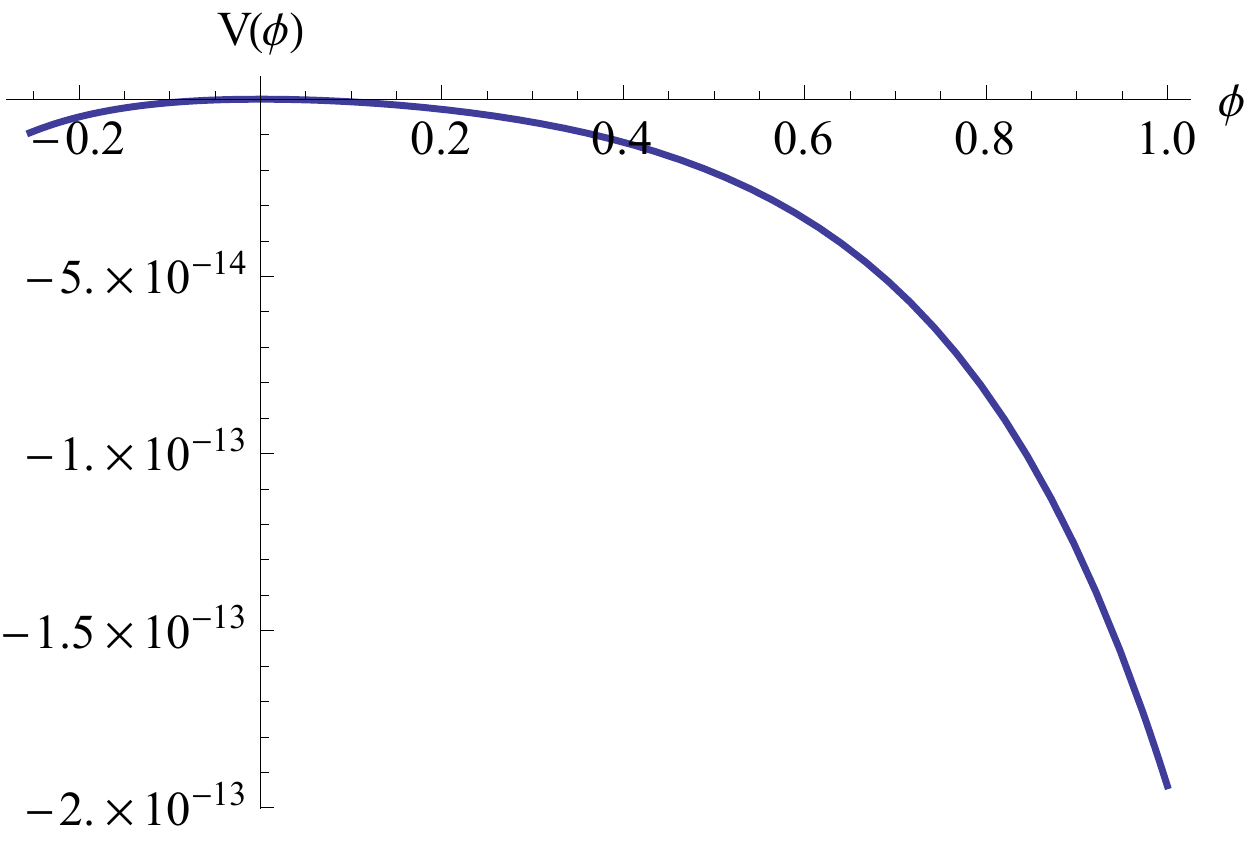}
\caption{The nature of the approximate unique Einstein frame potential
  for $\beta=-10^{12}$ and $\gamma=\frac{2}{3}\beta^2$ in Planck units.This
  potential only reflects the true dynamics of cubic $f(R)$ theory
  when $R \to 0$.}
\label{fig:auefp}
\end{center}
\end{figure}  

In the case of cubic gravity $$RF-f= \beta R^2 + 2\gamma R^3\,.$$
There is an interesting approximation using which one may avoid the
problems arising out of the multiple possible scalar potentials in the
Einstein frame. If one just neglects the cubic term in $R$ while
inverting the relation between $\phi$ and $R$ then this gives rise to
the approximate but unique result
$$R \sim \frac{1}{2\beta}\left(e^{\sqrt{2\kappa/3}\,\phi}-1\right)\,.$$ Now
one can write an unique scalar potential for cubic gravity in the
Einstein frame as
\begin{eqnarray}
V(\phi) \sim \frac{1}{2\kappa}\left[
\frac{1}{4\beta}e^{-\sqrt{2\kappa/3}\,2\phi}\left(e^{\sqrt{2\kappa/3}\,\phi}-1\right)^2 
+\frac{\gamma}{4\beta^3}e^{-\sqrt{2\kappa/3}\,2\phi}
\left(e^{\sqrt{2\kappa/3}\,\phi}-1\right)^3\right]\,,
\label{bc}
\end{eqnarray}
as obtained in Ref.~\cite{Barrow:1988xh}. From the form of the above
potential one can verify that $V(\phi) \le 0$ for all $\phi$ as
$\beta<0$ and $\gamma>0$. The above potential has only one zero for
finite values of $\phi$ and that zero is at $\phi=0$. From now onwards
we will call this Einstein frame potential as the approximate unique
Einstein frame potential. This approximate unique Einstein frame
potential is defined for all values of $\phi$ although it truly
reflects reality only in the neighborhood of $\phi=0$. The nature of
the approximate unique Einstein frame potential for cubic gravity is
shown in Fig.~\ref{fig:auefp}. In this potential only the region near
$R=0$ is relevant and later we will show that this region also admits
a cosmological bouncing solution in the corresponding Jordan frame.
  
On the other hand if no approximations are made then one has two
scalar potentials $V_{1}(\phi)$ and $V_2(\phi)$ in the Einstein frame
\begin{figure}
\begin{center}
\includegraphics[scale=.75]{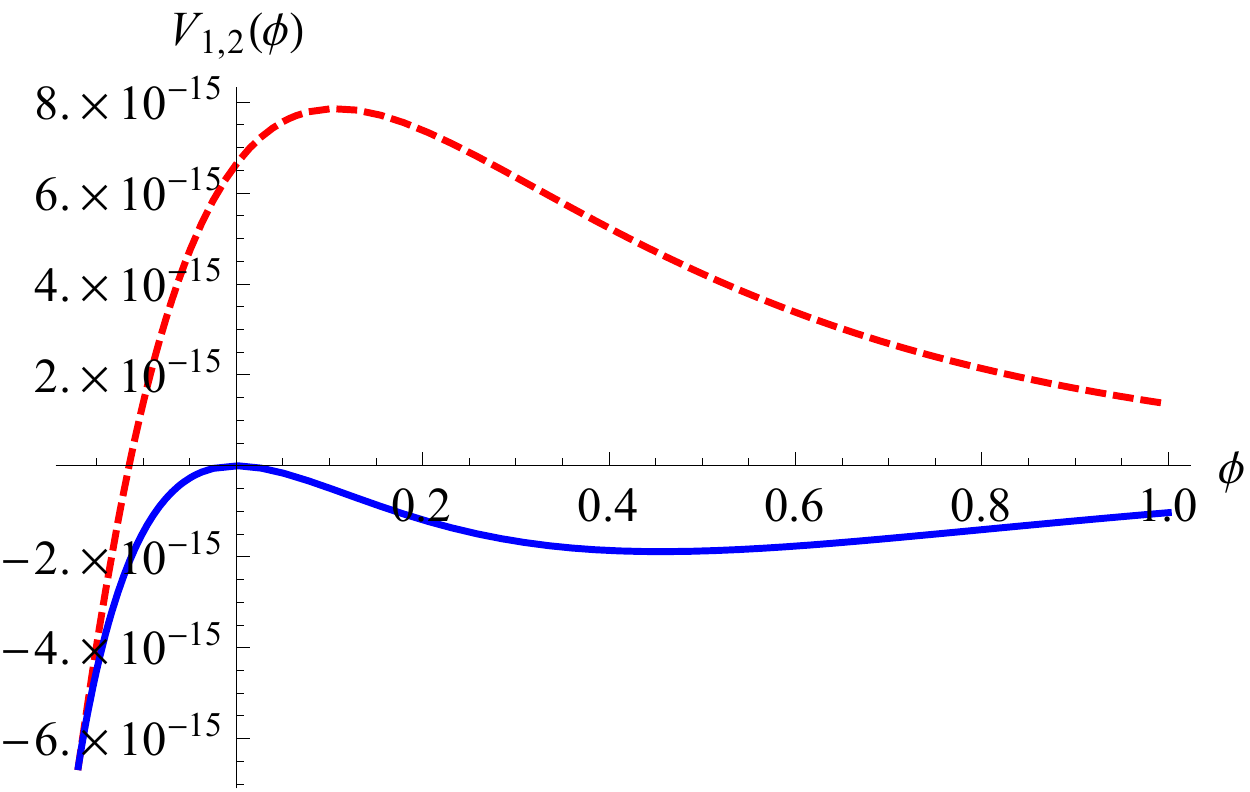}
\caption{The two branches of the exact Einstein frame scalar field
  potential for $\beta=-10^{12},\,\,\gamma=\frac{2}{3}\beta^{2}$ in Planck
  units. The solid curve is $V_{2}(\phi)$ and the dashed curve is
  $V_{1}(\phi)$. Here $dV_{1,2}/d\phi\longrightarrow\infty$ at
  $\phi=\phi_{c}$. The potentials do not exist for $\phi<\phi_{c}$.}
\label{fig:v12}
\end{center}
\end{figure} 
corresponding to the cubic gravity Lagrangian in Jordan frame. The
potentials can be written as
\begin{eqnarray}
V_{1,2}(\phi) &=& \frac{e^{-\sqrt{2\kappa/3}\,2\phi}}{2\kappa}
\left[\left(\frac{\beta}{3\gamma}\right) \mp
\sqrt{\left(\frac{\beta}{3\gamma}\right)^2 - 
\frac{1}{3\gamma}\left(1-e^{\sqrt{2\kappa/3}\,\phi}\right)}
\right]^2\nonumber\\
&\times &\left[\frac{\beta}{3} \pm
2\gamma\sqrt{\left(\frac{\beta}{3\gamma}\right)^2 - 
\frac{1}{3\gamma}\left(1-e^{\sqrt{2\kappa/3}\,\phi}\right)}
\right]\,,
\label{v12}
\end{eqnarray}
where subscript 1 corresponds to the upper minus sign and 2 corresponds to
the lower plus sign on the right hand side. These potentials
correspond to the two branches of $R$ in the $R-\phi$ plane.
The above forms of the scalar
field potentials in the Einstein frame immediately shows that as soon as
\begin{eqnarray}
e^{\sqrt{2\kappa/3}\,\phi} < \left(1-\frac{\beta^2}{3\gamma}\right)\,,
\label{validc}
\end{eqnarray}
the Einstein frame description of cubic gravity ceases to exist as the
potentials turn out to be complex. From the expressions of
$V_{1,2}(\phi)$ it is clear that $V_2(\phi) \le 0$ for all values of
$\phi$, for the bouncing solution, where $\beta < 0$ and
$\gamma>0$. More over $V_2(\phi)$ has a zero at $\phi=0$. On the other
hand $V_1(\phi)$ can have both positive and negative values for the
above choice of parameters and it has a zero for $\phi \ne 0$. It can
also be noted that both $V_1(\phi)$ and $V_2(\phi)$ are zero as $\phi
\to \infty$.
  
As an exponential function cannot ever be negative so the term on the
right hand side of the inequality, in Eq.~(\ref{validc}), cannot be
negative if the inequality holds. If one chooses the parameters
$\beta$ and $\gamma$ in such a way that the
$\left(1-(\beta^2/3\gamma)\right)$ is negative then the inequality
never holds and the Einstein frame description of cubic gravity holds
for all values of $\phi$. If one wants to make the Einstein frame
description of cubic gravity valid for all values of $\phi$ then
another issue related to the conformal transformations become
problematic. The Einstein frame description of cubic gravity holds as
long as $F(R)>0$, otherwise one cannot define the scalar field
itself. More over if $F(R) \le 0$ then the original $f(R)$ theory of
gravity itself becomes unstable.  If one requires $F(R)>0$ then the
inequality in Eq.~(\ref{bgama}) has to be satisfied and in that case
the right hand side of the inequality in Eq.~(\ref{validc}) become
positive and consequently there arises a finite value of $\phi$ as
\begin{eqnarray}
\phi_c = \sqrt{\frac{3}{2\kappa}}\ln\left(1-\frac{\beta^2}{3\gamma}\right)\,,
\label{svalidc}
\end{eqnarray}
which sets the lower limit of the scalar field strength in the
Einstein frame. The conformal transformations will never be able to
generate any Einstein frame dynamics of cubic gravity if
$\phi<\phi_c$. At this value of $\phi$, which corresponds to 
\begin{eqnarray}
R_c =-\left(\frac{\beta}{3\gamma}\right)\,.
\label{rc}
\end{eqnarray}
one can verify that $F'(R)=0$ and as a consequence the Einstein frame
description of cubic gravity becomes ambiguous as discussed in
subsection \ref{instabfr}. The two potentials $V_1(\phi)$ and
$V_2(\phi)$ are in general different from each other and
$V_1(\phi)=V_2(\phi)$ only when $\phi=\phi_c$ and $\phi \to \infty$
i.e., at the two extremities of the range of $\phi$. The approximate
scalar potential $V(\phi)$, as given in Eq.~(\ref{bc}), approximately
equals $V_1(\phi)$ very close to $\phi=0$. For this choice one can see
that the Ricci scalar in the Jordan frame approaches zero as $\phi \to
0$ in the Einstein frame. Consequently the approximate potential, as
proposed in Ref.~\cite{Barrow:1988xh} only holds near the $R \to 0$
limit in the Jordan frame and for all other finite values of the Ricci
scalar one has to rely on the dynamics produced by one of the actual
potentials $V_{1,2}(\phi)$ in the Einstein frame. The nature of the
two potentials $V_1(\phi)$ and $V_2(\phi)$ for cubic $f(R)$ gravity
are shown in Fig.~\ref{fig:v12}. The plots show that the two
potentials meet at $\phi_c$ which is negative for the particular
choice of parameters $\beta$ and $\gamma$ in cubic gravity. The
potentials also merge at $\phi \to \infty$ which can be inferred from
the plots as the two potentials converge as $\phi$ increases in the
positive side. The figure shows that $V_2(\phi)$ is always negative
except at $\phi=0$ and $V_1(\phi)$ has a zero for a non-zero $\phi$.

The above arguments show that the Einstein frame description of cubic
gravity (defined in Jordan frame) is limited. On the other hand one
may partition the Einstein frame description of cubic gravity in two
regions dictated by $V_1(\phi)$ and $V_2(\phi)$ and apply the
conformal transformations safely but as soon as one reaches $\phi_c$,
where these two potentials merge with each other, the Einstein frame
description fails.  The next section shows that the Jordan frame
analysis of the cosmological system near $R=R_c$, where the Einstein
frame description fails, reveals an interesting feature of cubic
$f(R)$ theory.  Using the approximate potential, as given in
Eq.~(\ref{bc}), one never feels this constrained nature of cosmological
dynamics in the Einstein frame, as in this approximation the potential
is valid for all values of $\phi$, although it truly describes reality
only near $\phi=0$.
\subsection{The cosmological behavior of general polynomial $f(R)$
  theories very near to $R=R_c$ in the Jordan frame}
  \label{instability}

We have seen in the above discussions that the point $R=R_c$, where
$F'(R_c)=0$, has some special properties as far as conformal
transformations are concerned. In this subsection we will elaborate a
bit more about the specific nature of cosmological dynamics, in the
Jordan frame, when the system is near $R=R_c$. This analysis is
necessary as near $R=R_c$ the conformal transformations fail and
one cannot generate any Einstein frame analog of the Jordan frame
dynamics and consequently one has to rely only on the Jordan frame
dynamics of the system. For this analysis we will employ linear
perturbation theory near $R=R_c$ where the Hubble parameter is
$H_c$. As Eq.~(\ref{fried}) and Eq.~(\ref{2ndeqn}) are written purely
in terms of $R$, $H$ and $\rho$, for the spatially flat FLRW case, it is
enough to specify these parameters near $(R_c,H_c,\rho_c)$ point, in
the $R-H-\rho$ space, as
\begin{eqnarray}
R &=& R_{c}+\delta R\,,\nonumber\\
H &=& H_{c}+\delta H\,,\nonumber\\
\rho &=& \rho_c + \delta \rho 
\nonumber  
\end{eqnarray}
where $\delta R$ , $\delta H$ and $\delta \rho$ are small fluctuations
of $R$, $H$ and $\rho$ over $R_c$, $H_c$ and $\rho_c$ for any
arbitrary $f(R)$ theory.  To linear order in fluctuations one can
write
\begin{eqnarray}
f(R) = f(R_{c})+F(R_{c})\,\delta R\,,
\nonumber
\end{eqnarray}
near the $(R_c,H_c)$ point where $F(R_{c})=f'(R)|_{R=R_c}$. As we have
assumed $F'(R_{c})=0$ we have,
\begin{eqnarray}
F(R) &=& F(R_{c})\,,\nonumber\\
F'(R) &=& F''(R_{c})\,\delta R\,,\nonumber\\
F''(R) &=& F''(R_{c})+ F'''(R_{c})\,\delta R\,,
\nonumber
\end{eqnarray}
near the $(R_c,H_c)$ point. In this analysis we will assume that
$F(R)>0$ as this ensures stability of the $f(R)$ theory.  Using the
above results in Eq.~(\ref{fried}) and Eq.~(\ref{2ndeqn}) and
calculating order by order one obtains
\begin{eqnarray}
3H_{c}^{2} &=& \frac{\kappa}{F(R_{c})}\left[\rho_{c}+
  \frac{1}{2\kappa}\left\{R_c F(R_c)-f(R_c)\right\}\right]\,,
\label{lin1}\\
H_{c}\,\delta H &=& \frac{\kappa}{6F(R_{c})}\,\delta\rho\,,
\label{lin2}\\
\dot{H} &=& \dot{\delta H} = -\frac{\kappa (1+\omega)}
{2F(R_{c})}\,\delta\rho\,,
\label{lin4}
\end{eqnarray}
and 
\begin{eqnarray}
\rho_{c}(1+\omega) &=& 0\,.
\label{lin3}
\end{eqnarray}
The above set of equations reveals more information if one rewrites
Eq.~(\ref{lin1})in the following form:
\begin{eqnarray}
3H_{c}^{2} &=& \frac{\kappa}{F(R_{c})}\left[\rho_{c}+
  F^2(R_c)V(\phi_c)\right]\,,
\label{lin1m}
\end{eqnarray}
where in the above equation we have used the expression of $V(\phi)$,
as given in Eq.~(\ref{potphi}), an Einstein frame object deliberately
used for a specific reason.  For cosmological bouncing scenarios in
the absence of any hydrodynamic matter, in cubic gravity, one has
$V(\phi_c)<0$. As a consequence the above equation predicts that the
universe will never be able to reach the $R=R_c$ point after bounce as
at that particular cosmological epoch $H_c^2<0$. The discussion below
will show that not only for bouncing solutions in vacua, or any other
solutions with hydrodynamic matter, where Eq.~(\ref{lin1m}) gives
$H_c^2>0$, the evolving universe will never be able to reach the phase
when the Ricci scalar attains the value $R_c$.

To understand the general nature of cosmological evolution near
$R=R_c$ one has to concentrate on the physical content of
Eq.~(\ref{lin3}). This equation states that very near to the $(R_c,H_c,\rho_c)$
point either the universe is devoid of hydrodynamic matter or the
matter present must be acting like a fluid with an equation of
state $\omega=-1$. These two cases can be separately analyzed. First
we assume that the universe is filled up with matter where $\rho_c
\ne 0$ but $\omega=-1$. In this case one  attains a local de
Sitter universe around $(R_c,H_c,\rho_c)$ point as now $\dot{\delta
  H}=0$ from Eq.~(\ref{lin4}).  Much away from the $(R_c,H_c,\rho_c)$
point it is difficult to predict the nature of the Hubble parameter
but one can always predict uniquely the value of $\omega$ as it is a
constant in our analysis.  In this case it is observed that if the
universe at all comes near the $R=R_c$ point then that universe must
have a local de Sitter phase. If one starts to track the cosmological
evolution from some point away from the neighborhood of the
$(R_c,H_c,\rho_c)$ point one would see that the time required for such
a development is infinite.
\begin{figure}
\begin{center}
\includegraphics[scale=.75]{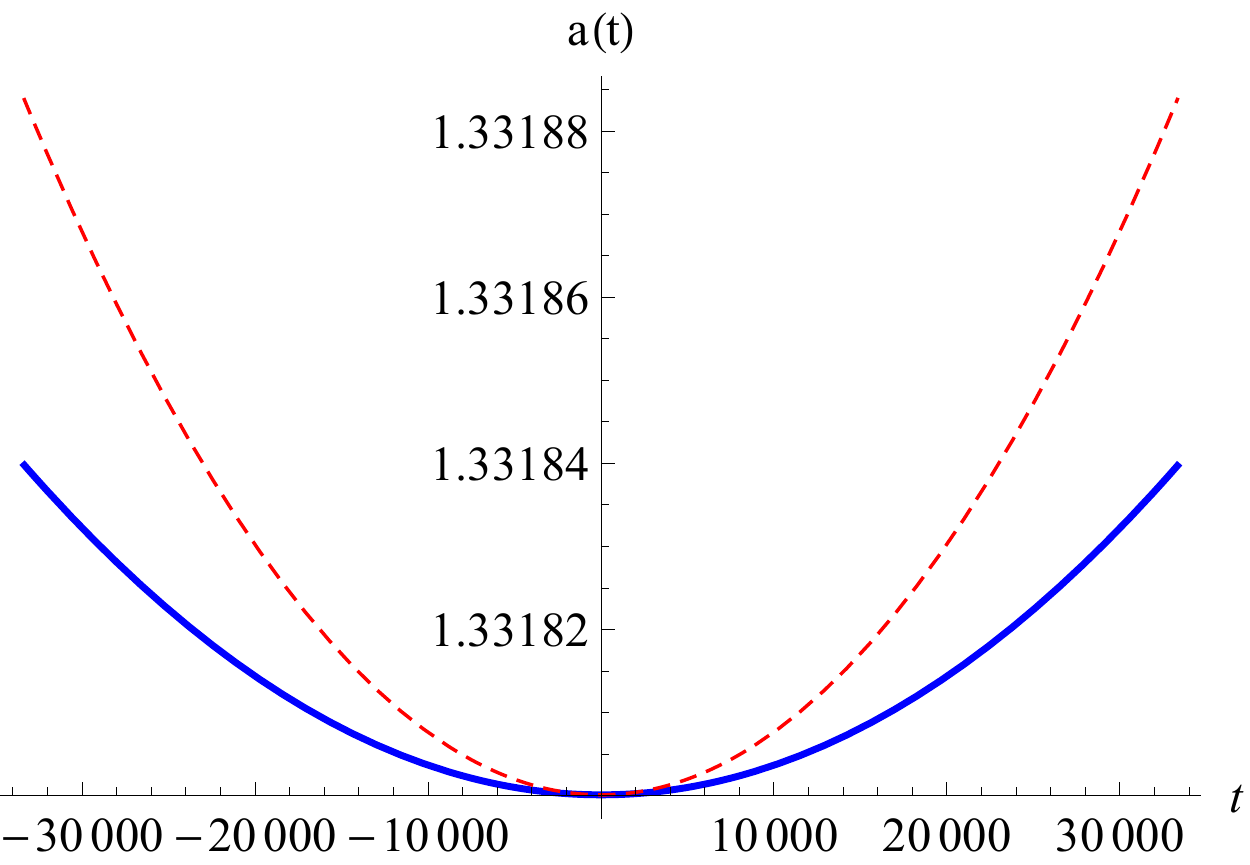}
\caption{Bounce with radiation background in branch $V_{2}(\phi)$,
  shown by the dashed curve, and in branch $V_{1}(\phi)$, shown by the
  solid curve. In these plots $\beta=-10^{12},\,\,\gamma=\frac{2}{3}\beta^{2}$ in
  Planck units.}
\label{fig:matt_bounce}
\end{center}
\end{figure}
The reason for this observation can be understood by the equation
governing time development which is given by
\begin{eqnarray}
t=\int_{0}^t \frac{dH}{\dot{H}}\,.
\nonumber
\end{eqnarray}
As the universe nears the $(R_c,H_c,\rho_c)$ point ${\dot H}\to 0$ and
consequently $t\to \infty$. This implies that for cubic gravity the
universe will take an infinite time to reach the $(R_c,H_c,\rho_c)$
point after bounce. 

If one particularly concentrates on a bouncing universe devoid of any
matter, we have shown previously that the $(R_c,H_c)$ point is never
attained as at that point $H_c$ turns out to be complex. In this
particular case in between the bouncing point where $H=0$ and
$\dot{H}>0$ and the $(R_c,H_c)$ point the universe has to pass through
a phase where ${\dot H}=0$ as because $H$ was increasing with time
after the bounce and must have decreased to zero at some time after
bounce so that $H^2$ becomes smaller than zero near the $(R_c,H_c)$
point. More over at the point where $\dot{H}=0$ one must have $H \ne
0$ so that $R \ne 0$. This fact can be understood from the nature of
the curve in Fig.~\ref{fig:phir}, where a bounce in absence of
hydrodynamic matter can happen for regions on the right hand side of
the minima of $\phi$ (in the $V_1(\phi)$ branch) and their $R \ne
0$. One can now apply the same logic around the point where $\dot{H}$
becomes zero and show that the universe will take an infinite time to
reach that point.

On the other hand if the universe had explicit hydrodynamic matter
during bounce, whose equation of state $\omega \ne -1$, then one can
see from Eq.~(\ref{lin3}) that the energy density becomes exactly zero
at the $(R_c,H_c,\rho_c)$ point. This can only happen when the
universe becomes infinite in volume (or the physical volume which
corresponds to a unit comoving volume diverges) such that the matter
energy density becomes exactly zero. If we assume that at a finite
cosmological time after the bounce the physical size of the expanding
universe to be finite, then the $(R_c,H_c,\rho_c)$ point can only be
reached by the universe after an infinite time. The density $\rho$, of
hydrodynamic matter, will in general decrease as the universe expands
after the bounce and tries to reach the $(R_c,H_c,\rho_c)$ point, but
it will never be exactly zero at any finite time.  Consequently in all
of the cases one can observe that the bouncing cosmological system,
guided by cubic gravity, will never reach the $(R_c,H_c,\rho_c)$ point
after a finite cosmological time from the bounce. The
$(R_c,H_c,\rho_c)$ point in a certain sense will always remain
inaccessible to the bouncing solutions in cubic gravity. The Einstein
frame description also precisely fails at these points. The discussion
of the nature of the cosmological system near the $(R_c,H_c,\rho_c)$
point is interesting and we have probed cosmological evolutions near
this point for bouncing solutions in cubic gravity. We expect our
results to be more general and valid for other forms of polynomial
gravity as well.
\begin{figure}
\begin{center}
\includegraphics[scale=.75]{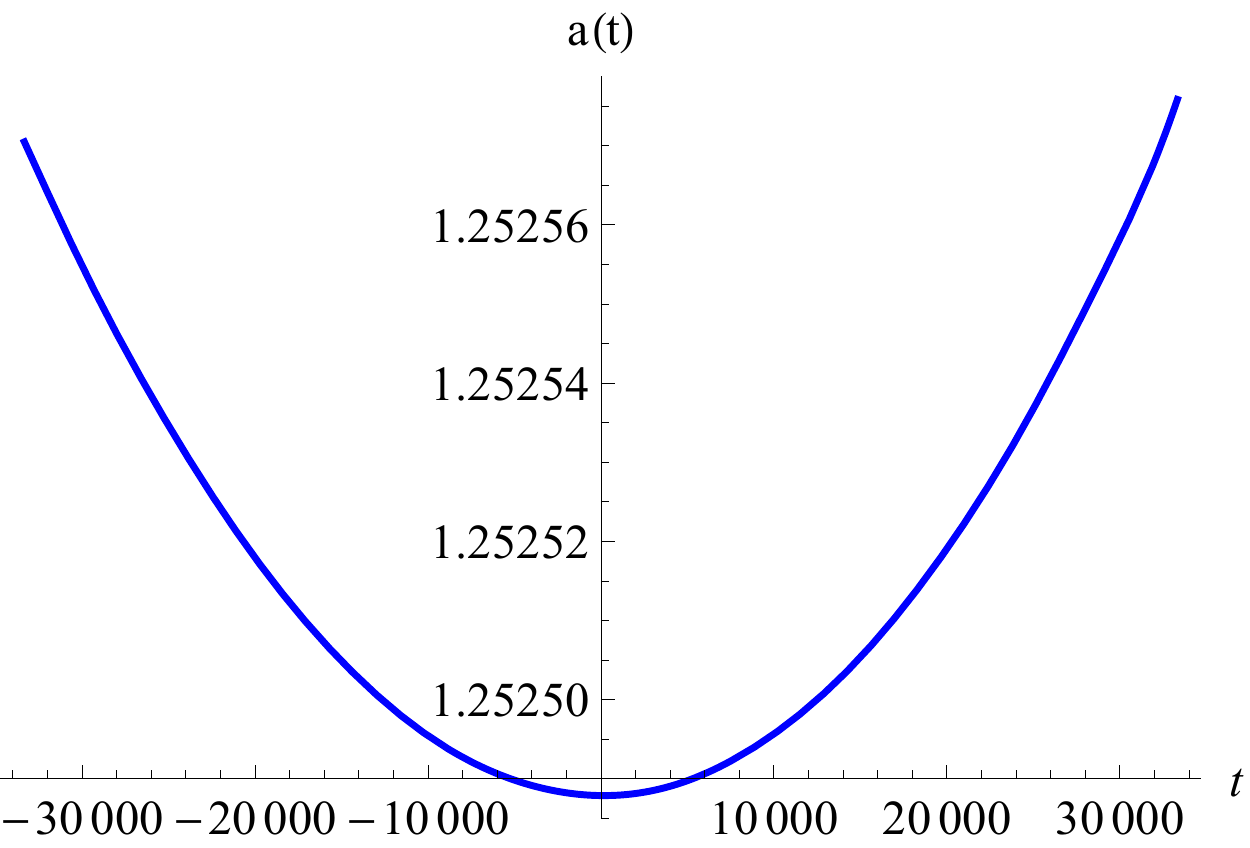}
\caption{Cosmological bounce in vacuum in branch $V_{1}(\phi)$ where 
$\beta=-10^{12},\,\,\gamma=\frac{2}{3}\beta^{2}$ in Planck units.}
\label{fig:matt_less_bounce}
\end{center}
\end{figure}
\section{Ways to choose between the three potentials in cubic gravity}
\label{3pot}

The natural question which arises in the case of polynomial $f(R)$
gravity is related to the proper choice of the potential in the
Einstein frame. The problem becomes non-trivial in the simplest of the
cases in cubic gravity where one can have three different potentials in
the Einstein frame which corresponds to a unique $f(R)$ in the Jordan
frame. The first amongst the three is the approximate unique Einstein
potential as given in Eq.~(\ref{bc}). This potential truly reflects
cubic gravity for small values of the Ricci scalar $R$.  The other two
potentials are $V_1(\phi)$ and $V_2(\phi)$ as given in
Eq.~(\ref{v12}). This potentials are in general valid for all values
of $R$ when either $f''(R)> 0$ or $f''(R)<0$.  In the following part of
this section we show that there can be interesting physics involved in
choosing the proper potential in the Einstein frame. We will show that
presence of matter during bounce can be an important factor which
helps us to choose amongst the three possible potentials in the
Einstein frame.

In cubic gravity one can have a cosmological bounce in the Jordan
frame in presence of matter. In this case one has to satisfy the
condition given in Eq.~(\ref{bpotc}).  Following the discussion after
Eq.~(\ref{bpotc}) in subsection \ref{enrgcond} it is seen that for a
successful cosmological bounce in the Jordan frame one must have
$V(\phi)<0$ in the Einstein frame.  As in this case a bounce always
happens when $V(\phi)<0$, we can infer that such a bouncing scenario
may be described by both $V_1(\phi)$, $V_2(\phi)$ or $V(\phi)$ as
given in Eq.~(\ref{bc}) in the Einstein frame, as all of these
potentials can have regions where $V(\phi)<0$. The behaviors of the
scale factor for the flat FLRW spacetime near cosmological bounces in
presence of radiation are shown in Fig.~\ref{fig:matt_bounce}. It is
seen that both the potentials are capable of producing cosmological
bounces in presence of matter.   
  
On the other hand if the universe is devoid of any hydrodynamic matter
then from Eq.(\ref{bpotc}) one can see that at the time of bounce
$V(\phi)=0$. In this case at the time of bounce, in general $\phi\ne
0$ in the Einstein frame. Consequently, pure curvature driven bounce
must happen through the potential $V_1(\phi)$ in the Einstein frame as
only $V_1(\phi)$ does have a zero for non-zero $\phi$. In this case if
${d\phi}/d\tilde{t}$ is non-zero when $V_1(\phi)=0$, the scalar field
climbs to positive values of $V_1(\phi)$ during the bouncing period.
The plot showing the evolution of the scale factor $a(t)$, for cubic
gravity, near a cosmological bounce in vacuum is shown in
Fig.~\ref{fig:matt_less_bounce}.
\begin{figure}
\begin{center}
\includegraphics[scale=1]{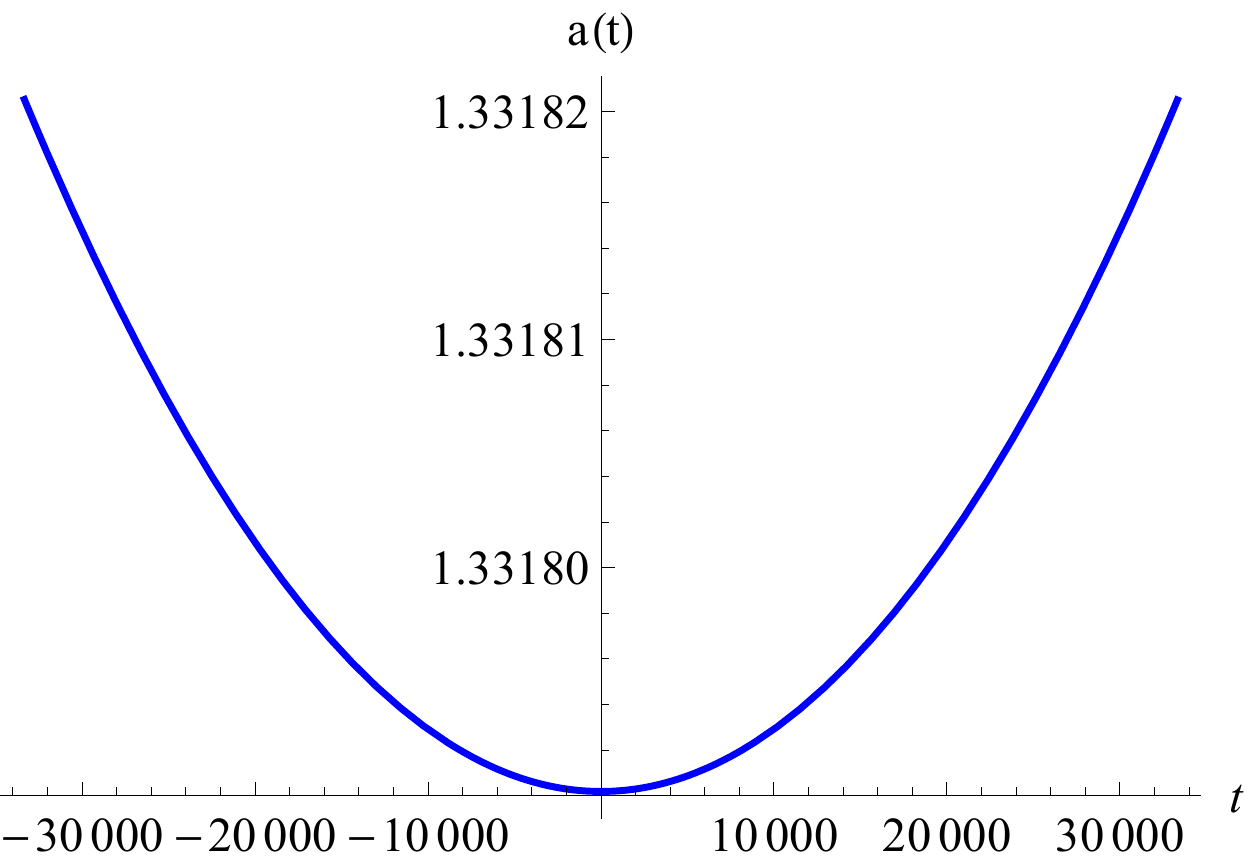}
\caption{Bounce with radiation background using the approximate
  unique Einstein frame potential with  
$\beta=-10^{12},\,\,\gamma=\frac{2}{3}\beta^{2}$ in Planck units.}
\label{fig:auefp_bounce}
\end{center}
\end{figure}

From the above description of bouncing phenomenon in the Jordan frame
we observe that although the approximate unique Einstein frame
potential can describe a cosmological bounce in presence of matter but
it has no information about pure curvature energy dominated
bounce. Consequently one can infer that the scope of the approximate
unique Einstein frame potential for describing cosmological bouncing
phenomenon in the Jordan frame is very limited. It is only a true
reflection of cubic gravity for very small values of the Ricci scalar
$R$ and more over it can only describe a matter induced bounce in the
Jordan frame. The nature of a cosmological bounce in cubic $f(R)$
theory where one employes the approximate unique Einstein frame
potential is shown in Fig.~\ref{fig:auefp_bounce}. This bounce can
only happen in presence of matter background and for our case we have
chosen the background hydrodynamic fluid to be comprised of radiation.

\subsection{Detailed description of the cosmological bounces in cubic
  $f(R)$ gravity}
\label{bounce_einstein}

 Discussions regarding pre-big bang cosmology which discusses about
 cosmological bounce was noted long ago, in 1992, in the
 Ref.~\cite{Gasperini:1992em}. It is well known that the cosmological
 behavior can be quite different in the Jordan frame and Einstein
 frame. In particular, it was mentioned in \cite{Paul:2014cxa} that a
 bounce in the Jordan frame is not usually associated with a
 corresponding bounce in the Einstein frame for flat FLRW
 cosmologies. In spite of these observations it is surprising that the
 Jordan frame bounce scenario can be understood from the Einstein
 frame scalar potential picture.

The evolution of the universe near the bouncing point in the Jordan
frame can be understood from the nature of Fig.~\ref{fig:phir} where
the minimum of $\phi$ occurs at $R=R_c$. On the right of $R_c$ in
Fig.~\ref{fig:phir} one sees that $\phi$ is an increasing function of
$R$. This can only happen for the branch where $\ln F(R)$ is an
increasing function of the Ricci scalar, and from the initial
discussions in section \ref{multiplicity}, this happens in the branch
$V_1(\phi)$. Consequently the branch $V_1(\phi)$ corresponds to that
part of cosmological dynamics in the Jordan frame where $R>R_{c}$.  In
this branch $R$ ranges from $R=R_c$ to $R=\infty$.  The other
potential branch $V_2(\phi)$ corresponds to the region $R<R_c$ in
Fig.~\ref{fig:phir} where $\phi$ is a decreasing function of the Ricci
scalar. This branch does have the embryo of Dolgov-Kawasaki like
instabilities when the Ricci scalar becomes small. On the other hand
as because in this branch the Ricci scalar can come close to zero, the
original $f(R)$ theory may cross-over to conventional GR in the
neighborhood of $R=0$.  From this discussion one can trace the
behavior of cosmological bounce in the Jordan frame if one follows the
behavior of the scalar field $\phi$ in the Einstein frame.

Let us first discuss about the particular potential branch $V_1(\phi)$
in the Einstein frame, which can support cosmological bounces in
presence or absence of hydrodynamic matter. First we discuss about
cosmological bounces in the absence of any hydrodynamic matter.
Bounce in the absence of matter in the Jordan frame occurs when the
scalar potential in the Einstein frame vanishes, as discussed
previously. In this branch, at the bouncing time $\tilde{t}=0$, the
value of the scalar field $\phi=\phi_0$ when $V_1(\phi_{0})=0$. In
this branch, there are several possible lines of development of the
cosmological system. These possibilities depend on the ``velocity''
($d\phi/d\tilde{t}$ or $\phi'$) of the scalar field at the time of
bounce. The contraction phase can start from $\phi<\phi_{0}$, cross
that point (Jordan frame bounce), rise up to a maximum along the
potential $V_1(\phi)$ and come down again. If the scalar field has
enough initial kinetic energy it can also cross over the maxima of
$V_1(\phi)$ and then slowly roll towards $\phi\rightarrow\infty$. If
it chooses to come down again it crosses $\phi=\phi_{0}$ again, but
this does not correspond to a bounce, which can be understood by a
careful look at the Einstein frame initial conditions. Then it goes
down towards $\phi_{c}$. While the universe moves towards $\phi_c$ in
the Einstein frame, the cosmological system in the Jordan frame moves
so that the Ricci scalar their becomes $R_c$. In doing so the universe
will require an infinite time in the Jordan frame. One can also
program bounce, in absence of matter, in such a way that the Einstein
frame scalar field starts from above $\phi>\phi_{0}$ during the
contracting phase, then the cosmological system crosses the bounce
point and goes down towards $\phi_{c}$. Cosmological bounce in
presence of hydrodynamic matter follows a similar course except for
the fact that at the time of bounce the scalar field assumes some
value $\phi=\phi_{0}$ where $V_{1}(\phi_{0})<0$. More over, one must
remember that bounce in the presence of matter can also occur in the
$V_2(\phi)$ branch of the potential.

Similar consideration holds for the potential branch $V_{2}(\phi)$ but
only for matter induced bounce. In this case $\phi$ assumes some value
$\phi=\phi_{0}$ at the time of bounce, such that
$V_{2}(\phi_{0})<0$. In this branch increasing or decreasing of $\phi$
correspond to decreasing or increasing of the scalar curvature $R$. 
The particular case of bounce in the $V_{2}(\phi)$ branch, where $\phi$ starts
from some $\phi<\phi_{0}$ and evolves towards $\phi= 0$, is
interesting. In this case the contracting phase may proceed to the
bounce point $\phi=\phi_0$ and then crosses this point and goes
towards $\phi=0$. In this branch as $R\to 0$ as $\phi\to 0$, as
apparent from Fig \ref{fig:phir}, one can assume that the universe is
evolving to a radiation dominated phase where the dynamics is governed
by general relativity. The region near the $R\to 0$ can act as a
cross-over region from $f(R)$ theory to general relativistic dynamics.

Before we end this discussion we present the various values of the
scalar field and its time derivative used in the Einstein frame to
numerically solve the equations dictating the bouncing phenomena. The
plot in Fig.~\ref{fig:matt_less_bounce} shows a cosmological bounce in
absence of hydrodynamic matter where
$\phi_0=-.11,\,\,\phi_0^{\prime}=10^{-6}$.\footnote{This implies
  $\phi_0=-.05M_{P}=-.05\times 10^{19}\,{\rm GeV}$;
  $\phi_0^{\prime}=-10^{-6}M_{p}^{2}=-10^{32}\,{\rm GeV}^{2}$.} Note that out
of these two conditions only $\phi^{\prime}$ can be chosen
suitably. The value of $\phi$ at bounce is already specified by the relation
in Eq.~(\ref{rbn}) which gives,
\begin{eqnarray}
\phi_{0}=\sqrt{\frac{3}{2\kappa}}\ln (1-\frac{\alpha^{2}}{4\beta})\,.
\label{phibn}
\end{eqnarray}
In this case $\phi$ starts from some point where $V_1(\phi)<0$ (see
Fig.~\ref{fig:v12}), crosses $\phi_{0}$, rises up to a maximum along
the potential and then comes down again.

The plots in Fig.~\ref{fig:matt_bounce} and
Fig.~\ref{fig:auefp_bounce} shows cosmological bounces in presence of
radiation in both the branches $V_{1}(\phi)$, $V_{2}(\phi)$ and in the
approximate unique Einstein frame potential for the same initial
conditions $\phi_0=-.13,\,\,\phi_0^{\prime}=10^{-8}$. Unlike the
matterless case, here both the initial conditions can be chosen
suitably, with the only condition that the scalar field potential at
bounce has to be negative. In these cases $\phi$ starts from some
point $\phi<-.13$, crosses the point $\phi=\phi_0$, goes up to a
maximum along the potential and then comes down again. The important
thing to be noticed in the above mentioned bouncing phenomena, in
presence of matter, is that the same conditions at the time of bounce
produces three different kinds of time dependence of the scale factor
$a(t)$ in the Jordan frame. Consequently these bounces are distinctly
different and shows the richness of polynomial $f(R)$ gravity.
\section{Conclusion}

In this paper we have extended our analysis on cosmological bounces,
for the spatially flat FLRW solutions, in higher derivative $f(R)$
theories of gravity. In our earlier work \cite{Paul:2014cxa} we
proposed a scheme by which one can study the phenomena of cosmological
bounces, using spatially flat FLRW solutions, in $f(R)$ theories using
only the equations of general relativity. The method relied on the
validity of the conformal transformations which connected cosmological
dynamics in the Jordan frame, where the dynamics is governed by $f(R)$
theory, and the auxiliary Einstein frame which is used to do the
calculations related to the problem. Once the problem is solved in the
Einstein frame one can map the solutions to the Jordan frame and get
the bounce dynamics. This program can be flawlessly applied for the
case of quadratic gravity where $f(R)=R+\alpha R^2$ where $\alpha$ is
some constant negative real number. In the case of quadratic gravity
it turns out that there is only one unique Einstein frame description
of the bouncing phenomena. It was immediately noticed that the methods
used in the earlier publication were not adequate when one uses higher
order polynomials as cubic $f(R)$. The Einstein frame picture of cubic
gravity and other higher order polynomial gravity does not have a
one-to-one correspondence with the Jordan frame description of the
cosmological evolution and consequently one does not know a priory
which picture in the Einstein frame corresponds to the actual picture
in the Jordan frame. In the present work we have proposed a consistent
method using which one can map the cosmological bouncing problem from
the Einstein frame to the Jordan frame. The present work shows that a
proper understanding of the conformal transformations yields
interesting properties of the cosmological bounces in the Jordan
frame. Unlike the case of quadratic gravity, in higher order
polynomial gravity there are multiple ways in which cosmological
bounces can occur. In the case of quadratic gravity only bounce in
presence of hydrodynamic matter was allowed where as in cubic gravity
one can also have cosmological bounces in absence of any form of
hydrodynamic matter.

The issue of energy conditions was briefly touched in the present
work. We have shown that one can always produce a cosmological bounce
in the Jordan frame without violating the weak energy condition and
the null energy condition in the Einstein frame. The strong energy
condition and the dominant energy conditions may be violated near the
bouncing point in the Einstein frame. The interesting thing about the
Einstein frame analysis, of the bouncing problem, is that all the
energy conditions are valid at ${\tilde t}=0$, which corresponds to
the actual bouncing time in the Jordan frame.  For a cosmological
bounce the energy conditions in the Jordan frame are violated but one
must have to remember that the energy conditions are not that clearly
and uniquely formulated in the Jordan frame. As a result one may not
be very serious about the energy conditions in the Jordan frame.

In this work we have introduced the techniques to understand the
origin of multiple scalar potentials in the Einstein frame for cubic
or higher order $f(R)$ gravity. We have explicitly worked out the
cosmological bounces arising out of the various potential branches in
the Einstein frame. Our discussion also includes the approximate
Einstein frame potential which one can use to study cosmological
behavior near $R=0$ as given in \cite{Barrow:1988xh}. As soon one
includes the multiple potentials in the Einstein frame the richness of
the cosmological scenario becomes apparent. In the penultimate section
we have presented the ansatz following which one may map from the
Einstein frame picture to the Jordan frame picture without any confusion. 

The other important observation which is reported in the present work
is related to the cosmological behavior of the universe as it evolves
from a bounce to a state where the Ricci scalar attains a value
$R=R_c$ where $R_c$ is the root of the equation $f''(R)=0$. Here the
primes denote derivatives with respect to the Ricci scalar. It was
noticed that the conformal transformations between the Jordan frame
and the Einstein frame breaks down precisely at the point $R=R_c$ for
cubic gravity. In the present work we show that the state of the
universe when $R=R_c$ can never be connected to the bouncing universe
in a finite time. In other words the flat FLRW universe, described by
cubic $f(R)$ theory in the Jordan frame, would take an infinite time
to attain $R=R_c$ after the cosmological bounce. The roots of
$f''(R)=0$ bifurcates the region of cosmological existence. We have
presented our results for cubic gravity but we hope that our results
are general in nature and can hold true for higher order polynomial
gravity. This particular issue requires further research. More over
the points $R=R_c$ are interesting, as out of the the two branches
emanating from this point in Fig.~\ref{fig:phir}, the one on the right
of $R=R_c$ has $f''(R)>0$ and the one on the left of $R=R_c$ has
$f''(R)<0$. On the left branch one may expect Dolgov-Kawasaki like
instability in the low curvature limit of the theory. But as this
branch also includes the point $R=0$ one has the freedom for a
cross-over to more conventional GR dynamics as the Ricci scalar
becomes small in magnitude. Ideally the Dolgov-Kawasaki instability
cannot be applied to the very early universe, in $f(R)$ theory, as at
those times the curvature may be very high and consequently one can
always use the left branch to model cosmological bounces happening in
the very early universe. 

Lastly, we have utilized all the conceptual understanding about cubic
gravity to actually solve the cosmological bounce problem. The results
are given in Fig.~\ref{fig:matt_bounce},
Fig.~\ref{fig:matt_less_bounce} and Fig.~\ref{fig:auefp_bounce}. These
results are obtained by solving the problem in the Einstein frame and
then mapping the solutions back to the Jordan frame. These solutions
show various bounce possibilities in the Jordan frame for the case of
cubic gravity. We expect the possibilities of various kind of
cosmological bounces will increase with the order of polynomial
gravity as more and more Einstein frame scalar potentials will emerge
in the conformally connected Einstein frame. In this paper we do not
present the theory of cosmological perturbations as our main aim was
to analyze the complicated and rich background cosmological evolution
in polynomial $f(R)$ theories. We will present the perturbations on
this background in a forthcoming publication. Although, the use of the
Einstein frame as an auxiliary frame to calculate the background
bouncing dynamics was purely a technical choice it turned out that
this technical choice yields much valued information about the actual
Jordan frame cosmological evolution if one can disentangle the multi
valued correspondence between the conformal frames. The technique
evolved in the present work can be valuable as one can use pure
general relativistic techniques to solve a problem in a higher
derivative $f(R)$ theory of gravity.  Although the methods discussed
in this article were particularly aimed to tackle the problem of
bouncing cosmologies in the Jordan frame, we firmly believe that many
of the techniques developed in this article can also be applied to
understand the general nature of $f(R)$ theories.

\end{document}